\documentclass[12pt,preprint]{aastex}
\shorttitle{}
\shortauthors{Nesvorn\'y et al.}

\begin{document}
\baselineskip 19.pt

\title{OSSOS IXX: Testing Early Solar System Dynamical Models\\using OSSOS Centaur Detections}

\author{David Nesvorn\'y$^1$, David Vokrouhlick\'y$^2$, Alan S. Stern$^1$, Bj\"orn Davidsson$^3$, Michele T. Bannister$^4$,
Kathryn Volk$^5$, Ying-Tung Chen$^6$, Brett J. Gladman$^7$, J. J. Kavelaars$^{8,9}$, Jean-Marc Petit$^{10}$,
Stephen D. J. Gwyn$^{8}$, Mike Alexandersen$^{6}$}
\affil{(1) Department of Space Studies, Southwest Research Institute,\\
1050 Walnut St., Suite 300, Boulder, CO, 80302, United States}
\affil{(2) Institute of Astronomy, Charles University,\\ 
V Hole\v{s}ovi\v{c}k\'ach 2, CZ--18000 Prague 8, Czech Republic}
\affil{(3) Jet Propulsion Laboratory,\\ M/S 183-401, 4800 Oak Grove Drive, 
Pasadena, CA 91109, United States}
\affil{(4) Astrophysics Research Centre, Queen's University Belfast,\\ Belfast BT7 1NN, United Kingdom}
\affil{(5) Lunar and Planetary Laboratory, University of Arizona,\\ 1629 E University Blvd, Tucson, AZ 85721, United States}
\affil{(6) Institute of Astronomy and Astrophysics, Academia Sinica,\\ 11F of AS/NTU Astronomy-Mathematics Building, 
Nr. 1 Roosevelt Rd., Sec. 4, Taipei 10617, Taiwan, R.O.C.}
\affil{(7) Department of Physics and Astronomy, University of British Columbia,\\ Vancouver, BC, Canada}
\affil{(8) NRC-Herzberg Astronomy and Astrophysics, National Research Council of Canada,\\ 5071 West Saanich Rd, 
Victoria, British Columbia V9E 2E7, Canada}
\affil{(9) Department of Physics and Astronomy, University of Victoria,\\ Elliott Building, 3800 Finnerty Rd, 
Victoria, BC V8P 5C2, Canada}
\affil{(10) Institut UTINAM UMR6213, CNRS, Univ. Bourgogne Franche-Comt\'e,\\ OSU Theta F25000 Besan\c{c}on, France} 
\begin{abstract} 
\normalsize We use published models of the early Solar System evolution 
with a slow, long-range and grainy migration of Neptune to predict the orbital 
element distributions and the number of modern-day Centaurs. The model distributions are biased by the OSSOS 
survey simulator and compared with the OSSOS Centaur detections. We find an excellent match 
to the observed orbital distribution, including the wide range of orbital inclinations 
which was the most troublesome characteristic to fit in previous models. 
A dynamical model, in which the original population of outer disk planetesimals was 
calibrated from Jupiter Trojans, is used to predict that OSSOS should detect $11\pm4$ 
Centaurs with semimajor axis $a<30$ au, perihelion distance $q>7.5$ au and diameter $D>10$ km 
(absolute magnitude $H_r<13.7$ for a 6\% albedo). This is consistent with 15 actual 
OSSOS Centaur detections with $H_r<13.7$. The population of Centaurs is estimated to be 
$21,000\pm8,000$ for $D>10$ km. The inner scattered disk at $50<a<200$ au should contain $(2.0\pm0.8) \times 
10^7$ $D>10$ km bodies and the Oort cloud should contain $(5.0 \pm 1.9) \times 10^8$ $D>10$ km comets. 
Population estimates for different diameter cutoffs can be obtained from the size distribution of Jupiter 
Trojans ($N(>\!\!D) \propto D^{-2.1}$ for $5<D<100$ km). We discuss model predictions for the Large 
Synoptic Survey Telescope observations of Centaurs.    
\end{abstract}

\keywords{}

\section{Introduction}

The Outer Solar System Origins Survey (OSSOS) is a wide-field imaging program that detected 838 
outer Solar System objects: a complete OSSOS database was recently released in Bannister et al. 
(2018). This can be compared, for example, to only 169 detections by the Canada-France Ecliptic 
Plane Survey (CFEPS) program (Petit et al. 2011). The orbits of OSSOS discoveries reveal new and 
complex detail in the distribution of Kuiper belt objects (KBOs). The OSSOS team has also developed a 
survey simulator, providing a straightforward way to account for OSSOS biases (Lawler et al. 2018a). 
The OSSOS database and simulator can be used to test different models of the early evolution of 
the outer Solar System.

The dynamical evolution of the early Solar System was reviewed in Nesvorn\'y (2018). Here we consider 
a class of models with slow, long-range and grainy migration of Neptune, because these models were
the most successful in reproducing the orbital distribution of KBOs (e.g., 
Nesvorn\'y \& Vokrouhlick\'y 2016; see Hahn \& Malhotra 2005, Levison et al. 2008, Nesvorn\'y 2015 
for related models). In brief, the outer planets are assumed to start in a 
resonant chain with Neptune initially at $\simeq$22-24 au. A massive outer planetesimal disk is 
placed from outside of Neptune's initial orbit to $\sim$30 au. The disk is dispersed during 
Neptune's migration with small fractions of the initial population of planetesimals ending on 
dynamically hot orbits in the present-day Kuiper belt.

The original outer disk is thought to have mass 15-20 $M_\oplus$ (Nesvorn\'y \& Morbidelli 2012), 
where $M_\oplus$ is the Earth mass, and a size distribution similar to that of today's observed 
Jupiter Trojans (Morbidelli et al. 2009). The suggested relation to Jupiter Trojans hinges on a capture 
model from Nesvorn\'y et al. (2013) (also see Morbidelli et al. 2005). Specifically, the Jupiter Trojan 
capture probability found in Nesvorn\'y et al. (2013) is $5\times10^{-7}$ for each outer disk planetesimal. 
There are 25 Jupiter Trojans with diameters $D>100$ km, which implies that the outer planetesimal disk 
contained $5\times10^7$ $D>100$-km planetesimals. Below 100 km, Jupiter Trojans have 
cumulative size distribution $N(>\!\!D) \propto D^\gamma$ with $\gamma = -2.1$ (Emery et al. 2015). 
From this we infer that the outer planetesimal disk contained $6\times10^9$ $D>10$-km 
planetesimals (Nesvorn\'y 2018).

The problem of calibration of the number and size distribution of disk planetesimals is important 
because it affects model inferences about various populations of small bodies in the Solar System. It 
has implications for our understanding of formational, collisional and dynamical processes 
in the early Solar System. The calibration from Jupiter Trojans, however, is not ideal because: (1) 
we cannot be entirely sure that the correct capture model has already been identified (see, e.g., Pirani 
et al. (2018) for a different capture model), and (2) the capture probability is somewhat uncertain 
even within the framework of our preferred model (Morbidelli et al. 2005, Nesvorn\'y et al. 2013). 
We thus feel compelled to consider other calibration methods. 

Centaurs detected by OSSOS provide an interesting constraint on the size distribution of the original 
planetesimal disk. It is well established that most Centaurs with $a<a_{\rm N}$, where
$a_{\rm N}$ is the semimajor axis of Neptune, and $q>7.5$ au (here we follow the definition 
of Gladman et al. 2008; Trojan and cometary orbits are excluded) evolved to their current orbits from the 
scattered disk (Duncan \& Levison 1997; DiSisto \& Brunini 2007; Volk \& Malhotra 2008, 2013). 
The scattered disk, in turn, 
formed from the original planetesimal disk when Neptune migrated into it and scattered planetesimals 
outward. A nice thing about this connection is that the implantation probability of bodies in the 
scattered disk and their subsequent evolution into the orbital realm of Centaurs are relatively 
insensitive to various model assumptions. In fact, all models proposed so far show that the current
population of the scattered disk should be 0.3-1.5\% of the original disk (e.g., Brasser \& 
Morbidelli 2013), with our preferred model consistently giving fractions near the lower end of 
this range (Nesvorn\'y et al. 2017). 

OSSOS detected 21 Centaurs (only tracked objects are used here) with absolute 
magnitudes ranging from $H_r=10.1$ to 16.1, which corresponds 
to $D=3$-50 km for a 6\% albedo (e.g., Bauer et al. 2013,
Duffard et al. 2014). This is ideal for the intended calibration because: (1) the detected sizes correspond 
to bodies that have not evolved collisionally {\it after} their implantation into the scattered disk 
(Nesvorn\'y \& Vokrouhlick\'y 2019); (2) they are well below the observed 'break' or 'divot' in the size distribution 
of large Kuiper belt objects (Bernstein et al. 2004, Shankman et al. 2013, Fraser et al. 2014), 
which simplifies modeling; and (3) the OSSOS-detected sample is large enough to constrain desirable 
quantities with reasonable confidence.

\section{Method}

We make use of the dynamical model from Nesvorn\'y \& Vokrouhlick\'y (2016). See this work
for the description of the integration method, planet migration, initial orbital distribution of 
disk planetesimals, and comparison of results with the orbital structure of the Kuiper belt.
In brief, the simulations track the orbits of the four giant planets (Jupiter to Neptune) and a large 
number of planetesimals. To set up an integration, Uranus and Neptune are placed inside of their 
current orbits and are migrated outward. The {\tt swift\_rmvs4} code, part of the {\it Swift} $N$-body 
integration package (Levison \& Duncan 1994), is used to follow the orbits of planets and (massless) 
planetesimals. The code was modified to include artificial forces that mimic the radial migration and 
damping of planetary orbits. These forces are parametrized by an exponential e-folding timescale, 
$\tau$ (Nesvorn\'y \& Vokrouhlick\'y 2016). 

The migration histories of planets were informed by our best models of planetary migration/instability 
(Nesvorn\'y \& Morbidelli 2012). We already demonstrated that these models provide the right framework 
to explain the orbital structure of the Kuiper belt and are also consistent with other properties 
of the Solar System (see Nesvorn\'y 2018 for a review). According to these models, Neptune's migration 
can be divided into two stages separated by a brief episode of dynamical instability (jumping Neptune 
model). Before the instability (Stage 1), Neptune migrates on a circular orbit. Neptune's eccentricity 
becomes excited during the instability and is subsequently damped by a gravitational interaction with 
disk planetesimals (Stage 2). Here we produced two different models corresponding to two different 
migration histories, which we refer to as the ``s10/30'' and ``s30/100'' cases (see Table~1).

The original planetesimal disk, from just outside Neptune's initial orbit to $\sim$30 au, is
assumed to be massive ($M_{\rm disk}=15$-20~$M_\oplus$; Nesvorn\'y 2018). Each simulation includes 
one million disk planetesimals. Such a fine resolution is needed to obtain good statistics for populations 
implanted into the Kuiper belt. The initial eccentricities and inclinations of disk particles are set 
according to the Rayleigh distribution. The disk particles are assumed massless, such that their gravity 
does not interfere with the migration/damping routines. 

The simulations tracked the orbital evolution of planets and planetesimals from the onset of Neptune's 
migration to the present time. To improve statistics 
for Centaurs, the orbits reaching $a<30$ au during the last 1 Gyr in our simulations were 
cloned 100 times. The cloning was accomplished by introducing a small (random) change $\delta V$ of the velocity 
vector ($\delta V/V \sim 10^{-5}$) of a particle when it first evolved to an orbit with $a<30$ au. The 
cloned orbits  were saved with a $10^4$ yr cadence producing a total of $1.7\times10^7$ Centaur orbits. They 
represent our dynamical model of the steady-state Centaur population. 

As for the size distribution, we want to test whether the original calibration inferred from Jupiter
Trojans gives the right number of Centaurs. We therefore adopt $N(>\!\!D) \propto D^\gamma$ with 
$\gamma=-2.1$ for $3<D<100$ km, and $N(>\!\!10\,{\rm km})=6\times10^9$ (see above).
We note that this slope is consistent with that surmised for the Kuiper belt from Pluto/Charon 
impact craters (Singer et al. 2019). It corresponds to $N(>\!\!H)\propto10^{\alpha H}$
with $\alpha = \gamma/5 = 0.42$, which is consistent with OSSOS observations of the scattered disk 
(Shankman et al. 2016, Lawler et al. 2018b). Whether the size distribution shows a break or divot 
near 100 km is irrelevant here because Centaurs detected by OSSOS have $D=3$-50 km. 
We use a fixed 6\% albedo (e.g., Grav et al. 2011, Duffard et al. 2014) to convert the size 
distribution into the magnitude distribution. The absolute magnitude distribution has to be specified 
in the r band, because all 21 OSSOS Centaurs were detected in r. 
 
The model distributions of Centaurs are used as an input for the OSSOS detection/track\-ing 
simulator, which was developed by the OSSOS team to aid the interpretation of their 
observations. The OSSOS simulator returns a sample of objects that would have been detected/tracked by 
the survey, accounting for flux biases, pointing history, rate cuts and object leakage (Lawler et 
al. 2018a). We use the OSSOS simulator output to determine whether the model results are consistent or 
inconsistent with the actual OSSOS detections. On one hand, predictions of a model that are inconsistent 
with the OSSOS detections can be used to rule out that model. On the other hand, our confidence in a 
specific model can be boosted if the model predictions turn out to be consistent with OSSOS. Note, 
however, that these arguments cannot be used to {\it prove} that a particular model is unique 
(simply because other, yet-to-be-tested models may fit data equally well).  
 
\section{Results}

\subsection{Orbit and Size Distributions} 

We elect to present our results in two steps. In the first step, we input the orbit and size 
distributions described above into the OSSOS simulator and let it generate 1000 (tracked) detections.
The resulting orbit and magnitude distributions are then compared to the actual OSSOS 
detections to test whether there is a good correspondence between the (biased) dynamical model 
and OSSOS observations. In the second step, we fix the number of Centaurs expected from our 
dynamical model by using the original calibration based on Jupiter Trojans (Nesvorn\'y 2018). We 
then run the OSSOS simulator to test how many Centaurs would be detected by OSSOS with the
original calibration.  

The biased model does a good job in reproducing the OSSOS detections (Figure \ref{fig1}). The K-S 
test gives 52\%, 93\%, 95\% and 60\% probabilities for the semimajor axis, perihelion distance,
inclination and magnitude distributions, respectively. The inclination distribution comparison 
is particularly satisfying because previous models with static planets (e.g., Figure~2 in Lawler 
et al. 2018b) were unable to account for the wide inclination distribution of Centaurs. 
In particular, Figure \ref{fig1} shows that the median {\it intrinsic} inclination of Centaurs is 
$\simeq$24$^\circ$, whereas the median inclination of {\it detected} Centaurs is $\simeq$14$^\circ$.
The wide inclination distribution is a consequence of slow migration of Neptune, which gives more opportunity 
to increase inclinations --by scattering encounters with Neptune-- before bodies are implanted into 
the Kuiper belt (Nesvorn\'y 2015).

The semimajor axis distribution of Centaurs detected by OSSOS shows a dip at 15-20 au, which is not 
reproduced in our model. The model, instead, shows a nearly linear trend with $a$. 
There is also a small difference between our model and OSSOS observations at the high end of the perihelion 
distance range. We find from the model that about 5\% of Centaurs detected by OSSOS should 
have $q>20$ au, whereas OSSOS did not detect any. Neither of these features is statistically 
significant, however.

More significantly, following the
definition of Gladman et al. (2008), we discarded Centaurs with $q<7.5$ au in Figure \ref{fig1}
(this includes four OSSOS objects with $q\sim5$ au). If the distributions shown in panel (b) of Figure 
\ref{fig1} are extended below $7.5$ au, we find that the model slightly over-predicts the number 
of detections for $q<7.5$ au. A more realistic model of this population would presumably need to  
account for a limited physical lifetime of bodies with low orbital perihelia (e.g., Levison \& Duncan 
1997). 

To obtain the $H_r$ distribution in panel (d) of Figure \ref{fig1}, we adopted a 6\% albedo (e.g., 
Grav et al. 2011, Bauer et al. 2013, Duffard et al. 2014) and only considered the magnitude range 
where OSSOS actually detected Centaurs (i.e., $H_r > 10$). If, instead, the magnitude distribution is 
extended to $H_r < 10$, the biased model indicates that $\simeq$15\% of Centaurs detected by OSSOS 
should have $H_r<10$. But OSSOS did not detect any Centaurs with $H_r<10$ (two Centaurs 
with $H_r \simeq 9$ and 9.5 were reported in the OSSOS ensemble catalog, but these come from 
other surveys; Petit et al. 2011, Alexandersen et al. 2016). In any case, the K-S 
test applied to the magnitude distribution gives a non-rejectable probability (30\%) 
even if the full magnitude range is used. Note that assuming a fixed albedo to convert between sizes 
and magnitudes is reasonable because all OSSOS Centaurs were found to be inactive (Cabral 
et al. 2019). 

\subsection{Absolute Calibration}

The second goal of this work is to test whether the number of Centaurs
detected by OSSOS is consistent with the original calibration from Jupiter Trojans. As we 
explained in Section 1, the number of $D>10$ km planetesimals in the original disk (i.e., before 
Neptune's migration) disk was estimated to be $6 \times 10^9$. Using this calibration 
and following planetesimals for 4.5 Gyr, we find that there should be 15,600 Centaurs with 
$a<30$ au and $D>10$ km. Diameter $D=10$ km corresponds to $H_r=13.7$ for a 6\% albedo. For 
reference, OSSOS detected 15 Centaurs with $H_r<13.7$ (detected objects with $q<7.5$ au
are excluded here).  

We therefore assume that there are presently 15,600 Centaurs with $a<30$ au and $H_r<13.7$, and 
run the OSSOS simulator on the model to determine the expected number of OSSOS detections. By 
repeating this test many times with random seeds we find that the OSSOS survey should detect 
$11 \pm 4$ (1$\sigma$ uncertainty) Centaurs (corresponding to the detection probability 
of $\simeq7\times10^{-4}$). This is to be compared to 15 actual detections (see above). We therefore see that 
the original calibration from Jupiter Trojans is consistent, at 1$\sigma$ level, with the number 
of Centaurs detected by OSSOS. This is an extraordinary result given that the dynamical models of the 
early evolution of the Solar System are often said to be limited in their predictive power.
The inferred size distribution of Centaurs is shown in Figure 2.

\subsection{Source Reservoirs}

We identified all objects that evolved onto Centaur orbits in our simulations
and tracked their orbits back in time to establish their orbital histories. All these objects 
started in the original planetesimal disk below 30 au (Section 2). Figure \ref{fig3} shows their orbits 1 Gyr 
ago when they resided in the trans-Neptunian region beyond 30 au. We find that 89\% of  
Centaurs had Kuiper belt/scattered disk orbits with $a<5000$ au and 11\% were in the Oort cloud
($a>5000$~au). For comparison, Nesvorn\'y et al. (2017) found that 95\% of ecliptic comets
(orbital period $P<20$ yr and the Tisserand parameter with respect to Jupiter $2<T_{\rm J}<3$).
evolved from orbits with $a<200$ au and 95\% of Halley-type comets ($20<P<200$ yr, $T_{\rm J}<2$) 
evolved from the Oort cloud.  

The source orbits of Centaurs show strong preference for $a<200$ au (85\% of the total). Of 
these, 31\% have $a<50$ au and 54\% have $50<a<200$ au. In this sense, the scattered disk beyond 
50 au is the main source of Centaurs, but the contribution from the classical/resonant Kuiper belt 
at $30<a<50$ au is also significant. For comparison, 20\% of ecliptic comets come from $30<a<50$ au and 
75\% from $50<a<200$ au (Nesvorn\'y et al. 2017). The preference for the scattered disk is 
therefore more pronounced for the ecliptic comets than for Centaurs. Also, 68\% (76\%) of 
Centaurs evolved from orbits with $q<35$ au ($q<36$ au) and $a<200$ au, which shows that 
the source orbits are typically at least marginally coupled to Neptune. This makes sense 
because the trans-Neptunian orbits with $q>36$ au are generally more stable and less often 
evolve to become planet crossing.
 
The fact that 11\% of Centaurs evolved from the Oort cloud in our simulations could explain
at least some the known very-high-inclination Centaurs (Gomes et al. 2015, Batygin \& Brown 2016).
OSSOS detected one Centaur with with $i=87^\circ$, $a=12.9$ au and $q=11.7$ au. This represents
a $\sim$6\% fraction of OSSOS Centaurs considered here. For comparison, orbits with $i>70^\circ$ 
represent only 0.6\% of our model Centaurs detected by the OSSOS simulator. The probability
of matching observations is thus roughly 1 in 10. Other sources of very-high-inclination Centaurs 
(Gomes et al. 2015, Batygin \& Brown 2016) may be needed at this level of significance. A more 
stringent constraint would be obtained with a larger ensemble of Centaurs.   

\subsection{2060 Chiron and 29P/Schwassmann–Wachmann 1}

Our dynamical model of Centaurs can be used to answer interesting questions about orbital 
evolution of specific objects. Here we illustrate these calculations for 2060 Chiron ($a=13.6$ au, 
$e=0.38$, $i=6.9^\circ$) and 29P/Schwassmann–Wachmann 1 (hereafter SW1; $a=5.99$ au, $e=0.044$,
$i=9.4^\circ$). The nearly circular orbit of SW1 with the perihelion distance at $q=a(1-e)=5.72$ au 
and the aphelion distance at $Q=a(1+e)=6.25$ au is unusual among Centaurs; it does not intersect
any planetary orbit. We may ask, for example, when SW1 evolved to its current orbit.

To answer this question, we select all simulated bodies with SW1-like orbits and calculate how 
long these bodies spent -on average and before arriving onto SW1-like orbits- with perihelion 
distance $q<6$ au and semimajor axis $a<7.5$ au. The answer is 38,000 yr. Using a typical
SW1 production rate, we estimate that $\sim$4\% of the SW1 mass would sublimate in 38,000 yr, 
thus eroding the SW1 diameter by $\sim$2 km (the SW1 diameter is estimated to be $\sim$50 km).

Another interesting question, with implications for the past activity of Chiron and SW1, is: 
``What is the probability that these objects had $q<3$ au (roughly the water ice sublimation 
radius) at any moment in the past?'' Here we select all simulated bodies that reached the 
present orbit of Chiron and compute the fraction of these bodies that had $q<3$~au before 
reaching Chiron's orbit. We find that the probability of Chiron having $q<3$ au in the past 
is only 7\%. The same calculation for SW1 gives 11\%. This shows that it is quite unlikely 
that any of these bodies experienced water-ice-sublimation-driven activity.

For comparison, all comets visited by spacecrafts had $q<1.6$ au when observations were 
made. In addition, from Nesvorn\'y et al. (2013) we estimate that $\sim$80-90\% of Jupiter 
Trojans had $q<3$ au {\it before} they were captured as Jupiter's co-orbitals. Morbidelli et al. 
(2005) quoted similarly high probabilities in their capture model (e.g., 68\% of Trojans
reached $q<2$ au before capture). This suggests that targets of the NASA Lucy mission were 
significantly more altered by solar heating (and water ice sublimation) than Chiron and SW1. 

\section{Discussion}

Recalibrating the number of planetesimals in the original disk from OSSOS Centaurs, we find that
there were $(8 \pm 3) \times 10^9$ planetesimals with $D>10$ km in the original outer disk. This implies 
only a minor adjustment of the population estimates given in Nesvorn\'y (2018). For example, 
the inner scattered disk at $50<a<200$ au should contain $(2.0 \pm 0.8) \times 10^7$ $D>10$ km bodies
and the Oort cloud should contain $(5.0 \pm 1.9) \times 10^8$ $D>10$ km comets. The error bars given 
above are standard 1$\sigma$ uncertainties that only take into account the number statistics of 
Centaurs detected by OSSOS. Additional uncertainties arise, for example, from the conversion between diameter and 
absolute magnitude. 

So far we discussed the results from the s30/100 model, where Neptune was assumed to have migrated on 
an $e$-folding timescale $\tau_1=30$ Myr before the instability and $\tau_2=100$ Myr after the 
instability. The preference for these long migration timescales is explained in Nesvorn\'y 
(2018). The results for s10/30 with $\tau_1=10$ Myr and $\tau_2=30$ Myr are similar, but we find 
that the inclination distribution of the biased s10/30 model is somewhat narrower (but 
non-rejectable). This is a consequence of shorter migration timescales in s10/30 that lead to 
somewhat smaller inclinations of orbits in the scattered disk. 

\begin{figure}
\epsscale{0.8}
\plotone{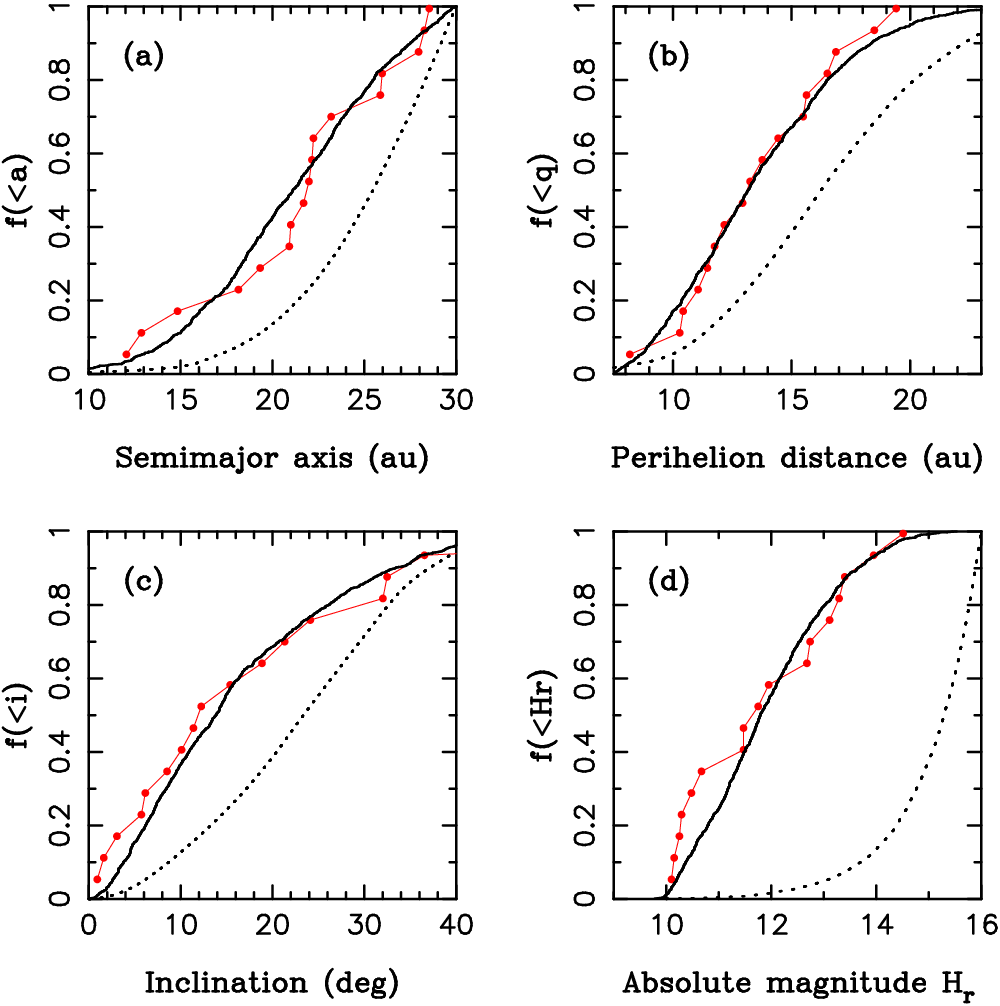}
\caption{Comparison of the biased model distributions (solid lines) with OSSOS detections of Centaurs
(connected red dots). Panels (a) to (d) show the semimajor axis, perihelion distance, inclination and absolute 
magnitude (fractional cumulative) distributions. The intrinsic distributions are shown as dotted lines. 
This result was obtained for the s30/100 model listed in Table 1. }
\label{fig1}
\end{figure}

\begin{figure}
\epsscale{0.6}
\plotone{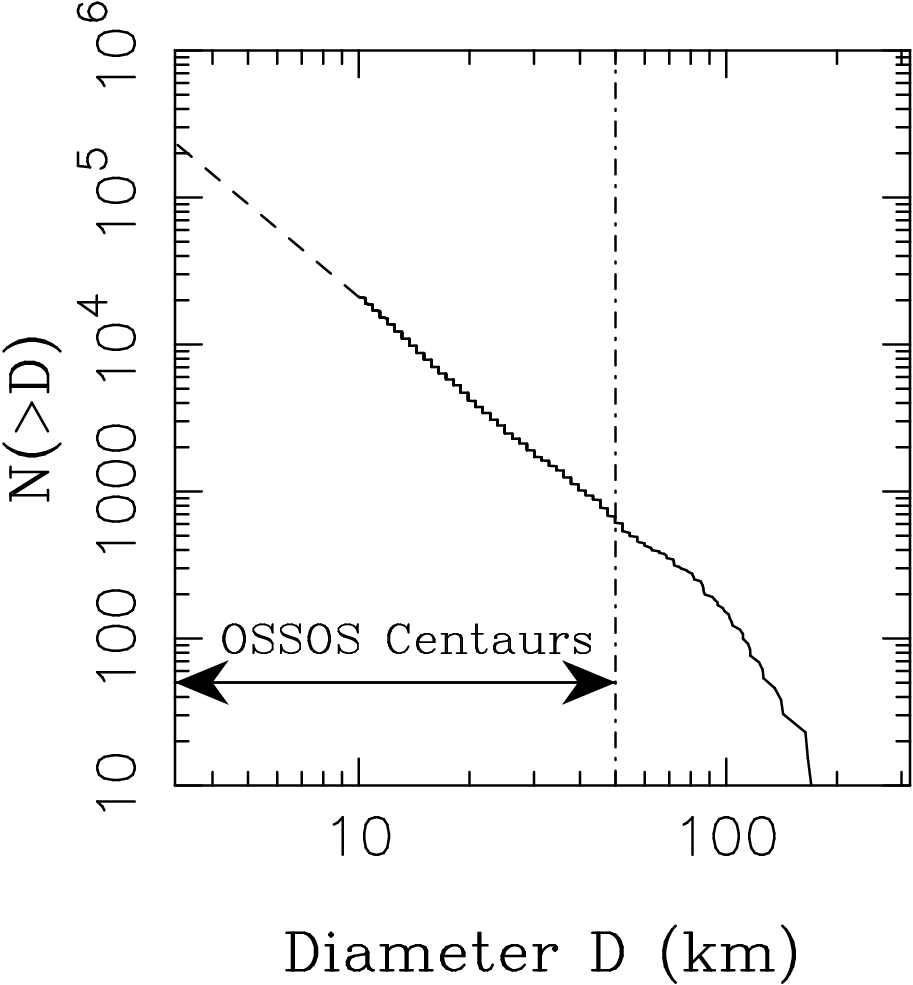}
\caption{The inferred size distribution of Centaurs. We extracted the size distribution of Jupiter Trojans 
from WISE (Grav et al. 2011), which is nearly complete down to $D=10$~km, and normalized it to 
21,000 Centaurs with $D>10$ km (see section 3.2). The dashed line for $D<10$ km shows an extrapolated 
distribution $N(>\!\!D)=21,000 \times (D/10\, {\rm km})^{-2.1}$. The size range of Centaurs detected 
by OSSOS is indicated by arrows.}
\label{fig2}
\end{figure}

\begin{figure}
\epsscale{0.6}
\plotone{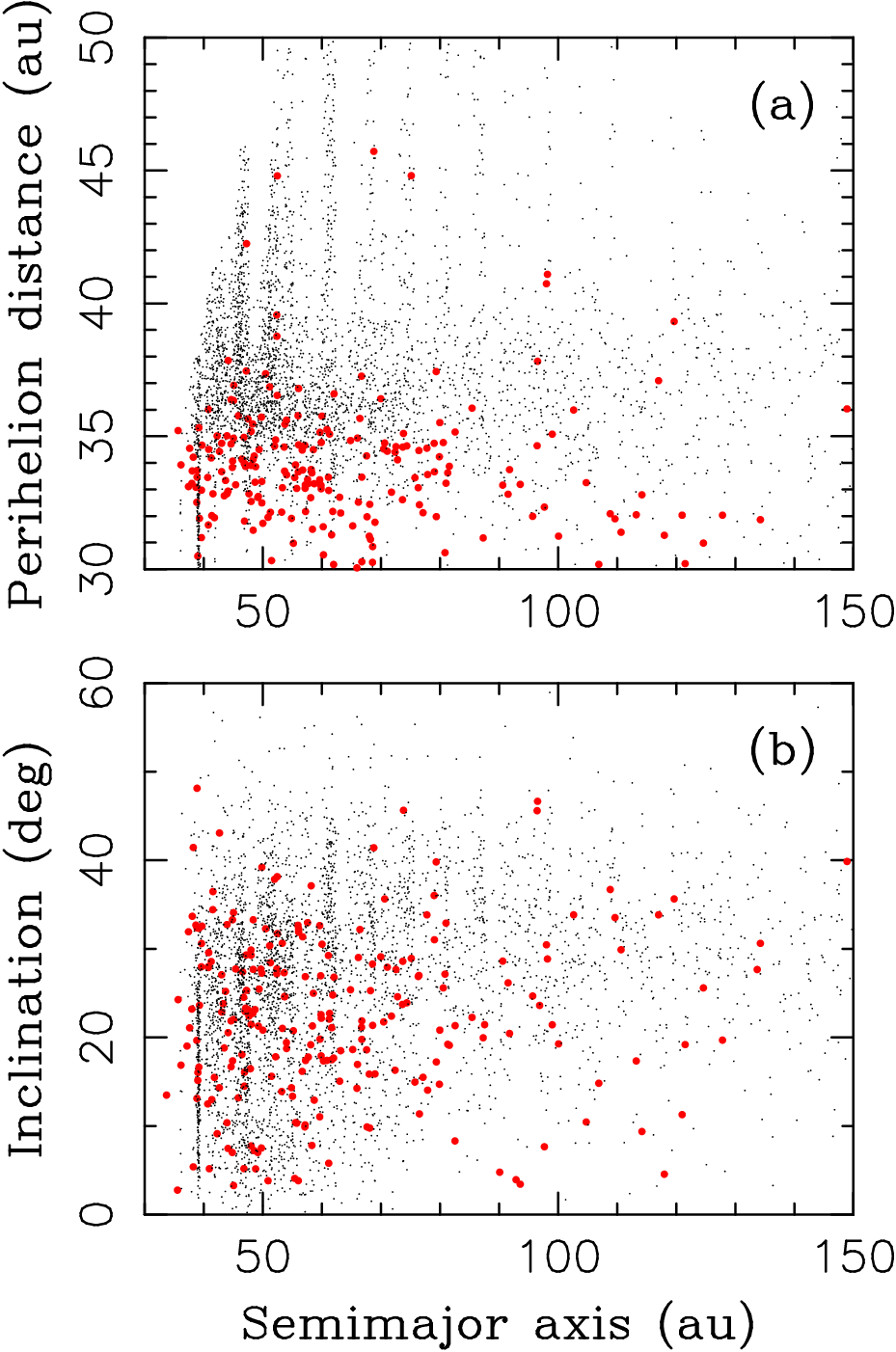}
\caption{The source reservoir of Centaurs. We identified all modern-day Centaurs in the s30/100 simulation 
and plotted their orbits 1 Gyr ago (red dots). The small black dots show the model distribution of 
trans-Neptunian objects 1 Gyr ago.}
\label{fig3}
\end{figure}

The intrinsic orbit (Figure \ref{fig1}) and size distributions (Figure \ref{fig2}) of Centaurs inferred 
here represent an interesting prediction for the Large Synoptic Survey Telescope (LSST) 
observations. We find that $\simeq$90\% and $\simeq$50\% of Centaurs should have $a>20$ au
and $a>25$ au, respectively. The median perihelion distance and median orbital inclination 
should be $\simeq$26 au and $\simeq$24$^\circ$. The population of Centaurs is 
estimated to be $21,000\pm8,000$ for $D>10$ km, $650\pm250$ for $D>50$ km and $150\pm60$ for $D>100$ 
km (estimates based on the size distribution shown in Figure \ref{fig2}; using $N(>\!\!D) = 21,000 (D/10)^{-2.1}$ would give
$720\pm280$ for $D>50$ km and $170\pm70$ for $D>100$ km). The estimate for $D>100$ km is consistent 
with $120^{+90}_{-60}$ Centaurs with $H_r<8.66$ ($D>100$ km for a 6\% albedo) from Lawler et al. (2018b).

\acknowledgements
The work of D.N. was supported by the NASA Emerging Worlds program. The work of D.V. was supported 
by the Czech Science Foundation (grant 18-06083S). M.T.B. appreciates support during OSSOS from UK 
STFC grant ST/L000709/1, the National Research Council of Canada, and the National Science and 
Engineering Research Council of Canada. K.V. acknowledges support from NASA (grants NNX15AH59G and 
80NSSC19K0785) and NSF (grant AST-1824869).

\clearpage
\begin{table}
\centering
{
\begin{tabular}{lrrrr}
\hline \hline
model        & $a_{{\rm N},0}$ & $\tau_1$ & $\tau_2$   & $N_{\rm Pluto}$  \\   
             & (au)       & (Myr)    & (Myr)      &            \\  
\hline
s10/30        & 24         & 10       & 30          & 2000       \\
s30/100       & 24         & 30       & 100         & 4000       \\
\hline \hline
\end{tabular}
}
\caption{A two stage migration of Neptune was adopted from Nesvorn\'y \& Vokrouhlick\'y (2016): 
$\tau_1$ and $\tau_2$ define the $e$-folding exponential migration timescales during these stages,
$a_{\rm N,0}$ denotes Neptune's initial semimajor axis,  
and $N_{\rm Pluto}$ is the assumed initial number of Pluto-mass objects in the massive disk below 30 au. 
Neptune's migration is grainy with these objects as needed to explain the observed proportion of 
resonant and non-resonant populations in the Kuiper 
belt.}
\end{table}

\clearpage

\clearpage

\clearpage

\end{document}